\documentclass{pasa}
\usepackage{graphicx,url,amssymb,amsmath,rotating,color,units,wasysym,epsfig,multirow,epstopdf,subcaption,xcolor}
\usepackage{subcaption}
\usepackage{tikz}
\usetikzlibrary{trees,positioning,shapes,shadows,arrows}
\tikzset{
  basic/.style  = {draw, text width=5cm, rectangle},
  root/.style   = {basic, rounded corners=3pt, thin, align=center, fill=white},
  level-2/.style = {basic, rounded corners=3pt, thin, align=center, fill=white, text width=2.2cm},
  level-3/.style = {basic, rounded corners=3pt, thin, align=center, fill=white, text width=1.7cm}
}
\usepackage{hyperref}

\title[Gravitational-wave bursts]{The DAWES review: Gravitational-wave burst astrophysics }

\author{
Jade Powell$^{1,2}$\thanks{dr.jade.powell@gmail.com},
Paul D. Lasky$^{3,4}$\thanks{paul.lasky@monash.edu},
\affil{$^1$Centre for Astrophysics and Supercomputing, Swinburne University of Technology, Hawthorn, VIC 3122, Australia}%
\affil{$^2$OzGrav: The ARC Centre of Excellence for Gravitational-Wave Discovery, Hawthorn, VIC 3122, Australia}%
\affil{$^3$Monash Astrophysics, School of Physics and Astronomy, Monash University, VIC 3800, Australia}%
\affil{$^4$OzGrav: The ARC Centre of Excellence for Gravitational-Wave Discovery, Clayton, VIC 3800, Australia}
}%

\jid{PASA}
\jyear{\the\year}

\usepackage{aas_macros}
\usepackage{hyperref} 
\hypersetup{colorlinks,citecolor=blue,linkcolor=blue,urlcolor=blue}

\begin{document}

\begin{frontmatter}
\maketitle

\begin{abstract}
Over a hundred gravitational-wave signals have now been detected from the mergers of black holes and neutron stars, but other sources of gravitational waves have not yet been discovered. Some of the most violent explosive events in the Universe are predicted to emit bursts of gravitational waves, and may result in the next big multi-messenger discovery. Gravitational-wave burst signals often have an unknown waveform shape, and unknown gravitational-wave energy, due to unknown or very complicated progenitor astrophysics. Potential sources of gravitational-wave bursts include core-collapse supernovae, cosmic strings, fast radio bursts, eccentric binary systems, and gravitational-wave memory. In this review, we discuss the astrophysical properties of the main predicted sources of gravitational-wave bursts, and the known features of their gravitational-wave emission. We summarise their future detection prospects, and discuss the challenges of searching for gravitational-wave burst signals and interpreting the astrophysics of the source. 
\end{abstract}

\end{frontmatter}

\section{Introduction}
What comes next in gravitational-wave astronomy? To date, all ground-based gravitational-wave observations have come from the coalescence and merger of neutron stars and black holes~\citep{gwtc1, gwtc2, gwtc3}. The increased sensitivity of the LIGO-Virgo-KAGRA~\citep{ligo, virgo, kagra} network of gravitational-wave observatories implies a significant and rapid increase in the volume of the Universe being surveyed, not only to compact-binary coalescences, but also to other types of sources. The positive detection of these other source types has the potential to vastly increase our understanding of  different aspects of astro- and fundamental physics, from neutron star and supernova physics, to more exotic aspects such as cosmic strings.

\begin{figure*}
\centering
\includegraphics[width=0.9\textwidth]{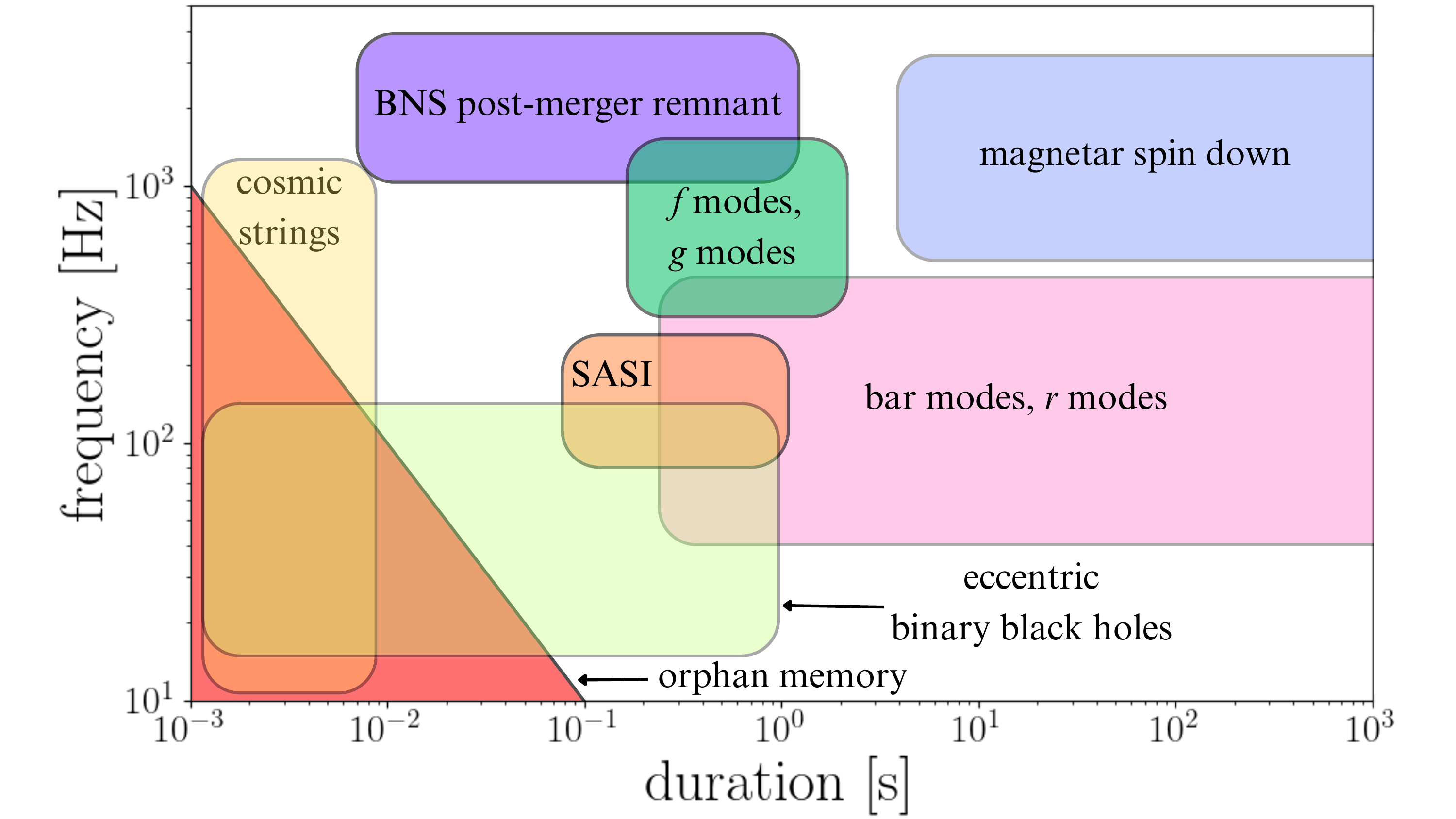}
\caption{The duration and frequency of different potential sources of gravitational-wave bursts. Bursts that last only a few ms include high mass eccentric binary black hole mergers, cosmic strings, and orphan memory. Sources that last up to a few thousand seconds include the spin down of millisecond magnetars. Gravitational waves from post mergers remnants and core-collapse supernovae can occur at frequencies above 1000\,Hz, including the different types of emission modes for core-collapse supernovae discussed in detail in Section~\ref{sec:ccsn}. }
\label{fig:frequencyduration}
\end{figure*}

Different gravitational-wave sources can be loosely divided into categories based on the duration of their signals. For example, `mountains' on rotating neutron stars create nearly monochromatic gravitational-wave signals that are persistent for months to millennia. Low-mass compact binary mergers last in the LIGO-Virgo-KAGRA band for many minutes, while relatively high mass binary black hole mergers ($\sim\unit[100]{M_\odot}$) are in band for less than tenths of a second. Other potential transient sources of gravitational waves also cover these timescales, from $\mathcal{O}(\unit[10-100]{ms})$ $f$-mode oscillations of neutron stars to $\mathcal{O}(\unit[]{day})$ long signals from fluid dynamics associated with pulsar glitch recovery. In this review, we focus on \textit{bursts} of gravitational waves with short durations from a few ms up to around 2000\,s. In Figure~\ref{fig:frequencyduration}, we show a schematic plot of the main sources of gravitational-wave bursts as a function of their gravitational-wave frequency on the vertical axis, and approximate duration on the horizontal axis. 

The majority of potential gravitational-wave burst sources are perfect targets for multi-messenger astronomy. A detection of gravitational waves from a burst source such as a core-collapse supernova (CCSN) may result in the first electromagnetic, neutrino and gravitational-wave joint detection, providing us with a comprehensive understanding of the underlying astrophysical processes \citep{mezzacappa_24}. A detection of gravitational waves from a fast radio burst (FRB) may provide us with some of the first insights into the nature of the source of these events \citep{lvk_frb_03a}. Some other examples of potential burst sources with electromagnetic counterparts are magnetar flares, gamma-ray bursts (GRBs), and pulsar glitches. Gravitational-wave bursts could also be emitted from an electromagnetically dark signal like a cosmic string, an eccentric binary black hole merger, or a previously unknown astrophysical object.

Making the first detection of gravitational-wave burst signals has many challenges. First and foremost, they are difficult to model. This may be due to complicated astrophysical processes in the progenitor, for example in CCSNe (see Section~\ref{sec:ccsn}), or because of a lack of knowledge of the nature of the source, such as FRBs (see Section~\ref{sec:frb}). A lack of understanding of the exact signal morphology means that typical template-based search techniques such as matched filtering cannot be used. Additional problems arise in that transient noise artefacts could potentially mimic these short-lived signals.

Current searches for gravitational-wave bursts tend to search for excess power that occurs coherently between multiple gravitational-wave detectors \citep[e.g.,][]{klimenko_16, lynch_17, sutton_10}, which mitigates transient noise artefacts parading as gravitational-wave signals. These searches make minimal assumptions about the morphology of the signal. Previous work has performed targeted searches for sources observed electromagnetically or through neutrinos \citep[e.g.,][]{lvk_magnetar_O3, lvk_frb_03a, icecube_23, GRB_O3b}. There are also all-sky and all-time searches to ensure that no gravitational-wave burst event is missed if there are no electromagnetic counterparts \citep{lvk_allsky_O3, lvk_long_O3}. To date, unmodelled burst searches have only made detections of binary black hole mergers, including the first gravitational-wave detection GW150914 \citep{gw150914}. They have not made any confident detections of any new types of gravitational-wave sources. 

In this review, we describe the various potential sources of gravitational-wave bursts. We discuss the probabilities of making a detection of these burst sources in the near future, or in next generation gravitational-wave observatories such as the Einstein Telescope \citep{et_11}, Cosmic Explorer \citep{evans_21}, or a dedicated kilohertz frequency observatory such as NEMO~\citep{ackley_20}. We describe how potential gravitational-wave signals can be used to interpret the astrophysical properties of the sources, and discuss what work still needs to be done for the community to be ready for the first detection from a new source. 
We begin in Section \ref{sec:ccsn} with a discussion of CCSNe. We discuss binary neutron star post merger remnants in Section \ref{sec:postmerge}, highly eccentric and hyperbolic compact binaries in Section \ref{sec:bbh}, FRBs in Section \ref{sec:frb}, pulsar glitches in Section \ref{sec:glitch}, magnetars in Section \ref{sec:magnetar}, GRBs in Section \ref{sec:grb}, orphan memory in Section \ref{sec:memory}, topological defects in Section \ref{sec:cosmo}, and unknown unknowns in Section \ref{sec:unknown}. We conclude in Section \ref{sec:conclusion}.

\section{Core-collapse Supernovae}
\label{sec:ccsn}

\begin{figure*}
\centering
\includegraphics[width=16cm]{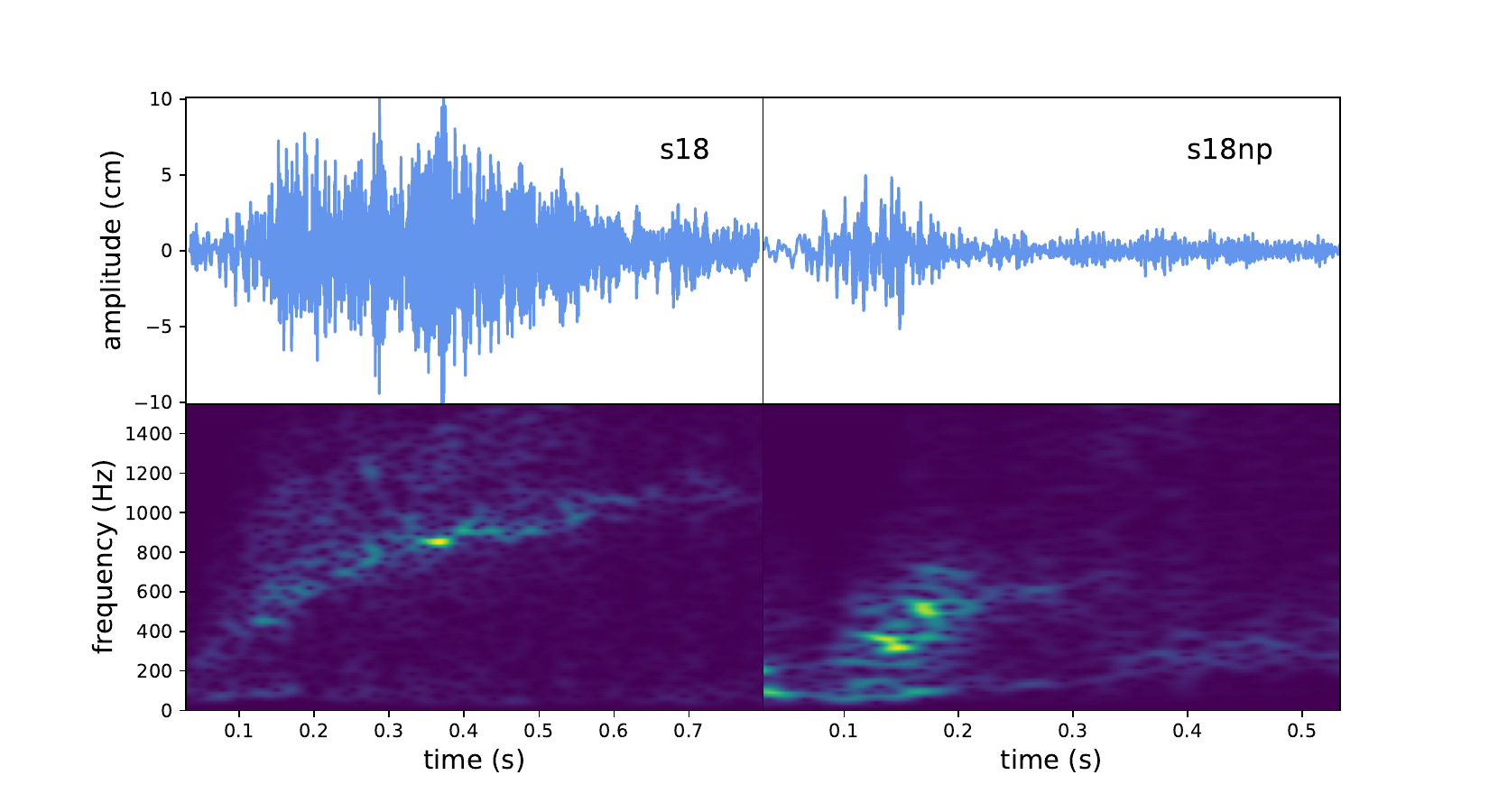}
\caption{Examples of typical gravitational-wave signals from core-collapse supernovae. Both models are $18\,\mathrm{M}_{\odot}$ stars from \citet{powell_19, Powell_20}, where model s18 (left panels) rapidly undergoes shock revival, and model s18np (right panels) fails to power a full supernova explosion. The top panels show the time series, and the bottom panels are spectrograms of the gravitational-wave signals. The main feature in the spectrograms is the high frequency g-mode, which has a frequency that is related to the properties of the proto-neutron star. The lack of shock revival in model s18np results in lower gravitational-wave amplitude and a longer duration low frequency mode due to the SASI.
 }
\label{fig:sn}
\end{figure*}


Core-collapse supernovae are the explosive deaths of stars more massive than $\approx8\,\mathrm{M}_{\odot}$. When a stellar core exceeds its effective Chandrasekhar mass it becomes gravitationally unstable and collapses until the core reaches nuclear densities. A shock wave is launched outwards from the core, loses energy, and stalls at a radius of $\sim150$\,km. The mechanism to revive the shock is not fully understood as electromagnetic observations cannot probe the inner regions of the star. However, gravitational waves and neutrinos are emitted directly from the core, and may provide us with the first direct probe of the mechanism that powers CCSN explosions \citep{mueller_17}.  

Predictions of gravitational waveforms for CCSNe come from numerical hydrodynamical simulations (see \citet{abdikamalov_20, mueller_20, mezzacappa_24} for recent reviews). However, CCSN simulations are extremely computationally challenging, and need to include multi-dimensions, accurate neutrino transport, stellar hydrodynamics, realistic equations of state, progenitor models from stellar evolution, general relativity, rotation and magnetic fields. There are some variations in gravitational-wave amplitude predictions between different simulation codes \citep{andresen_17, kuroda_17, radice_19, powell_19, mezzacappa_20}. However, there are many features of gravitational-wave signals from CCSNe that are considered to be robust, and have been seen in simulations made by many groups using different codes and different numerical set-ups.

The majority of CCSNe are thought to explode by the neutrino-driven explosion mechanism, where the shock wave is powered by absorbing some of the energy from the neutrinos \citep{janka_17}. More extreme supernovae are expected to explode by the magneto-rotational explosion mechanism, where the rotational energy of the star is transported into the shock wave by the magnetic fields \citep{reichert_23,mueller_24}. Several studies have shown how we can determine the explosion mechanism from the gravitational-wave signal using model selection techniques trained on waveforms from numerical simulations \citep{logue_12, powell_16, powell_17, saiz-perez_22, powell_23b}. Using currently available waveforms, different model selection algorithms are able to determine the explosion mechanism with high accuracy for gravitational-wave signals with a signal-to-noise ratio (SNR)$\gtrsim20$. 

There are currently a large number of waveforms available from three-dimensional simulations of neutrino-driven explosions \citep{andresen_17, kuroda_17, yakunin_17, oconnor_18, radice_19, powell_19, Powell_20, mezzacappa_20, pan_21}. On the other hand, there are only a small number of three-dimensional waveforms available for magneto-rotational explosions that extend beyond the core-bounce phase \citep{bugli_22, obergaulinger_22, powell_23, powell_24}. There are also significant differences in the magneto-rotational waveforms from different simulation codes \citep{varma_21}. More waveform development for magneto-rotational explosions is therefore needed to ensure we are ready to accurately determine the explosion mechanism for a real CCSN event. 

The main feature in CCSN gravitational-wave signals is the high frequency $g$/$f$-mode oscillations of the proto-neutron star, which usually has higher gravitational-wave energy than the other signal components. These modes have very clear features in signal spectrograms, as shown in Figure \ref{fig:sn}, where the gravitational-wave emission rises in frequency with time as the proto-neutron star radius shrinks and its mass increases. The mode typically starts around 100\,ms after the core-bounce time. The maximum gravitational-wave frequency of the mode can vary significantly between different simulations. For example, \citet{powell_19} see a maximum frequency of $\sim 1000$\,Hz in their neutrino-driven explosions, but \citet{pan_21} see a maximum frequency of $\sim 3000$\,Hz. 

Universal relations have been developed that describe the relationship between the gravitational-wave frequency and the mass and radius of the proto-neutron star \citep{torres_19, Sotani_21, sotani_24}, which has led to development of search and parameter estimation tools for this signal mode \citep{Powell_22, bizouard_21, Bruel_23}. These tools may enable us to learn proto-neutron star properties from a positive gravitational-wave detection. However, current universal relations ignore some aspects of the physics behind CCSNe, for example rotation and magnetic fields.  

In rotating models, there is a spike in the gravitational-wave time series at the time of core-bounce followed by smaller oscillations  \citep{dimmelmeier_08, abdikamalov_14, scheidegger_08}. If the star is non-rotating, then the gravitational waves at the core-bounce time will be effectively zero. The spike occurs when the equation of state stiffens and the proto-neutron star oscillates for a few ms. The bounce signal only occurs in the plus polarisation and the most optimal observer angle is from the equator of the star. So even if the star is rapidly rotating, there will be no core-bounce signal visible if we detect the gravitational waves from an observer angle near the poles. If the core-bounce signal is detected, it can be used to constrain properties of the star. For example, the amplitude of the bounce signal is related to the oblateness of the core that is determined by the rate of the rotation. Oscillations after the initial spike may tell us about the equation of state. The core-bounce signal is expected to be in the frequency range between 100-1000\,Hz. As this part of the signal is well understood, it is possible to perform template based searches and parameter estimation \citep{edwards_14, richers_17, pajkos_21, edwards_21, afle_21, pastor_23}.   

Another common feature found in CCSN gravitational-wave emission is a lower frequency mode due to the standing accretion shock instability~\citep[SASI;][]{blondin_03, blondin_06, foglizzo_07}. The SASI is an instability in the shock wave that can result in the proto-neutron star being excited from above producing pressure ($p$) modes. The SASI modes occur in the most sensitive frequency band of current gravitational-wave observatories ($\approx 200$\,Hz), which can aid in the signal's detectability. However, the gravitational-wave energy caused by the SASI is usually significantly lower than the energy in the $g$/$f$-mode. The gravitational-wave emission mode from the SASI usually increases in frequency with time up to the time of shock revival where the explosion prevents further growth of the SASI. Therefore, there is usually no SASI mode visible in models that quickly undergo shock revival. 

The gravitational-wave signal will also have a low-frequency component (below $\sim10$\,Hz) due to gravitational-wave memory. Matter motion post shock revival can result in a small memory component in the signal. However, asymmetric emission of neutrinos creates much stronger memory amplitudes, resulting in increased CCSN detection rates for gravitational waves in low-frequency detectors \citep{mukhopadhyay_21, vartanyan_23, powell_24}. Long-duration simulations are needed to fully understand the limits on the minimum frequency and maximum amplitude of this aspect of gravitational-wave emission. As this is currently not computationally feasible, some groups have tried to extend the duration of the gravitational-wave emission analytically \citep{richardson_22}. 

 A few studies have determined the detection prospects for CCSNe in current gravitational-wave observatories with unmodelled burst search algorithms \citep{gossan_16, Szczepanczyk_23}. 
For modern neutrino-driven CCSN explosion waveforms, the SNR needs to be approximately between 15 and 20 before the signal can be detected by an unmodelled burst search. This translates to Galactic detection distances, and puts the rates of CCSNe at a few per century for current ground based observatories \citep{taylor_14}. Magneto-rotational explosions may be detected at much greater distances, to beyond the Magellanic Clouds, but this does not add a significant improvement to the detection rates. 

\begin{figure*}
\centering
 \includegraphics[width=10.0cm]{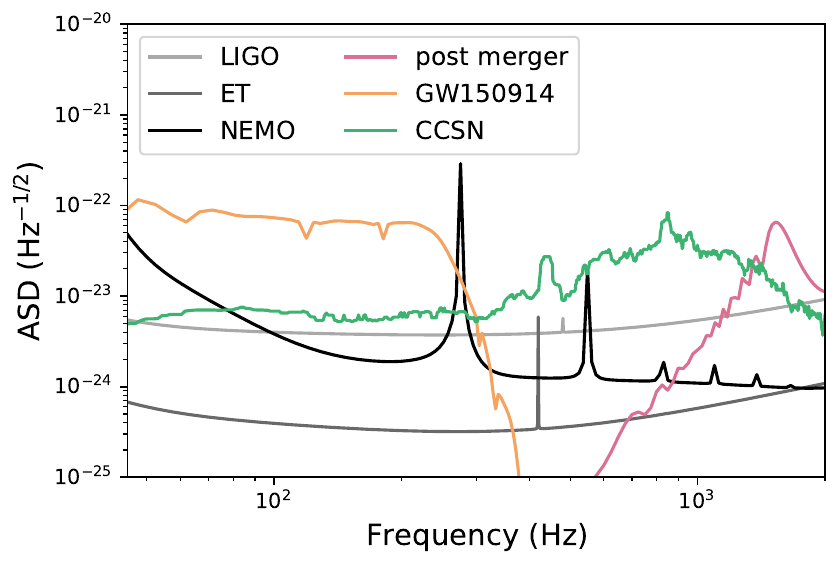}
\caption{The amplitude spectral density (ASD) of Advanced LIGO at design sensitivity, the Einstein Telescope (ET), the proposed NEMO observatory, a typical post merger signal, the binary black hole merger GW150914, and core-collapse supernova (CCSN) model s18 from \citet{powell_19} at a distance of 10\,kpc. Post merger signals occur in the higher end of the frequency range of the LIGO-Virgo-KAGRA observatories. Binary mergers typically occur at frequencies of a few hundred Hz. CCSN signals are broadband, with the majority of their amplitude above 500\,Hz.}
\label{fig:post_merg}
\end{figure*}

No gravitational waves from CCSNe have been detected during the first three Advanced LIGO-Virgo-KAGRA observing runs using unmodelled burst searches \citep{lvk_0102_SN, Szczepanczyk_23}.
In the next generation of ground based gravitational-wave observatories, the detection distances for CCSNe will vary from hundreds of kpc to a few Mpc, depending on the progenitor properties, potentially increasing the detection rates to as much as one per year \citep{powell_19, powell_24}. 
 
To enhance the prospects for further detections, further improvements need to be made to CCSN gravitational waveform predictions from numerical simulations, and also to the search algorithms for gravitational waves from CCSNe. The current searches are waveform agnostic, and need a relatively high SNR signal before the gravitational waves can be detected. If the known features of the gravitational-wave signals are incorporated into the search algorithms, then it may decrease the minimum SNR needed for the first CCSN detection. 

\section{Binary neutron star post merger remnants}
\label{sec:postmerge}

There are multiple potential outcomes for the remnants of binary neutron star mergers that primarily depend on the remnant's mass. If the remnant is below the so-called Tolman-Oppenheimer-Volkoff mass~$M_{\rm TOV}$~\citep{tolman_39, oppenheimer_39}, which is the maximum mass a non-rotating neutron star can attain without collapsing to a black hole, then an infinitely stable neutron star will be born from the collision. If the remnant mass is between $M_{\rm TOV}$ and $\chi M_{\rm TOV}$, where $1.3\lesssim\chi\lesssim1.6$ depending on the unknown nuclear equation of state~\citep[e.g.,][and references therein]{shibata_00,shibata_06,baiotti_17,agathos_20,bauswein_21}, then the remnant will be a meta-stable neutron star known as either a hyper- or supramassive neutron star. Hypermassive neutron stars, supported from collapse by differential rotation, are expected to survive for less than about one second~\citep[e.g.,][]{baumgarte_00,shapiro_00} before losing sufficient centrifugal support and collapsing to a black hole. Supramassive stars are supported from collapse by rigid rotation which is quenched on a longer timescale, and hence can survive for up to $\sim\unit[10^{4}]{s}$~\citep{ravi_14}. Remnants with mass above $\chi M_{\rm TOV}$ will immediately form a black hole as a direct result of the merger. For recent reviews of post-merger remnants, see~\citet{bernuzzi_20, sarin_21}.

Of the four stages mentioned above, the first three (stable, hypermassive, and supramassive) are all expected to emit large-amplitude gravitational waves that may be detectable in current- or next-generation gravitational-wave observatories for a short time ($\lesssim\unit[1]{s}$) after the merger. The amplitude of the emission is expected to be comparable, if not larger, than the amplitude of the inspiral strain. The characteristic frequency of emission is, on the other hand, almost predominantly in the kHz regime. Importantly, the specific frequencies emitted can be used to trace the nuclear equation of state~\citep{bauswein_12a,bauswein_12b,read_13}, implying the positive detection of post-merger remnants have the potential to significantly advance our understanding of nuclear physics in conditions inaccessible to terrestrial experiments.

While hypermassive remnants are expected to collapse on a timescale similar to that in which the gravitational-wave emission is expected to be visible, it looks unlikely that the collapse can be directly measured from the gravitational-wave signal even in third-generation observatories~\citep{easter_21,dani_24}. Long-lived supramassive and stable remnants, on the other hand, are expected to emit gravitational waves for significantly longer periods of time. These may be generated through magnetic-field-induced ellipticity of the remnant~\citep[e.g.,][]{cutler02,haskell08,dallosso09,lander20}, or stellar oscillation modes such as $r$ or bar modes~\citep[e.g.,][]{bondarescu09,corsi09}. However, the prospects of detection of these longer-lived signals has been vigorously debated in the literature~\citep[see][for a review]{sarin_21}, although many authors have tried searching for the long-lived remnant of the binary neutron star merger GW170817~\citep[e.g.,][]{abbott_17_gw170817_postmerger,abbott_19_gw170817_postmergerII,grace_24}. The long-lived nature of the emission in this case implies such gravitational-wave searches typically adopt methods from the community also searching for long-lived, nearly-monochromatic gravitational waves. As such, we declare these longer-lived signals out-of-scope for this review, and focus only on the short-lived signals in the immediate ($\lesssim\unit[1]{s}$) aftermath of the merger.

Astrophysically, what fraction of binary neutron star mergers create either a stable, hypermassive, or supramassive neutron star, and therefore emit the short-lived but loud gravitational-wave signals that we are discussing in this Section is currently unknown. 
There are two primary unknowns limiting our ability to answer this question: 1) the unknown TOV mass $M_{\rm TOV}$ and 2) the unknown mass distribution of binary neutron star merger progenitors. While the latter was believed to be reasonably well understood, the binary neutron star merger GW190425 has a total mass of $\unit[3.4^{+0.3}_{-0.1}]{M_\odot}$~\citep{abbott_20_gw190425}, which is inconsistent with the Galactic mass distribution of double neutron star systems~\citep{kiziltan_13,keitel_19,farrow_19}. We therefore wait patiently for more binary neutron star mergers to confidently determine this mass distribution for merging systems. 

So what became of the neutron stars in the landmark merger event GW170817 that was observed by the LIGO-Virgo-KAGRA collaboration~\citep{abbott_17_gw170817_detection,abbott_19_GW170817_properties} and  across the electromagnetic spectrum~\citep[e.g.,][]{abbott_17_gw170817_gwgrb,abbott_17_gw170817_multimessenger}? It is somewhat contentious to say that this is contentious. However, there are multiple authors who make definite claims that are in direct contradiction of one another about the nature of the remnant from e.g., the colour of the kilonova~\citep{margalit17,radice18,yu18}, the lack of early-time x-ray observations~\citep{piro19,ai20}, and late-time radio and x-ray observations~\citep{piro19,lin20,troja20}; for a review, see~\citet{sarin_21}. What is certain is the uncertainty about the long-term fate of the remnant, but it is highly likely that the remnant did not promptly form a black hole, and therefore emitted kHz gravitational waves. Despite multiple searches, reliable identification of gravitational waves from a post-merger remnant have not been identified, and estimates indicate the gravitational-wave observatories 
required at least an order of magnitude of additional sensitivity to make a detection
~\citep{abbott_17_gw170817_postmerger, krolak_23}.

\section{eccentric binaries and hyperbolic encounters}
\label{sec:bbh}

\begin{figure}
\includegraphics[width=\columnwidth]{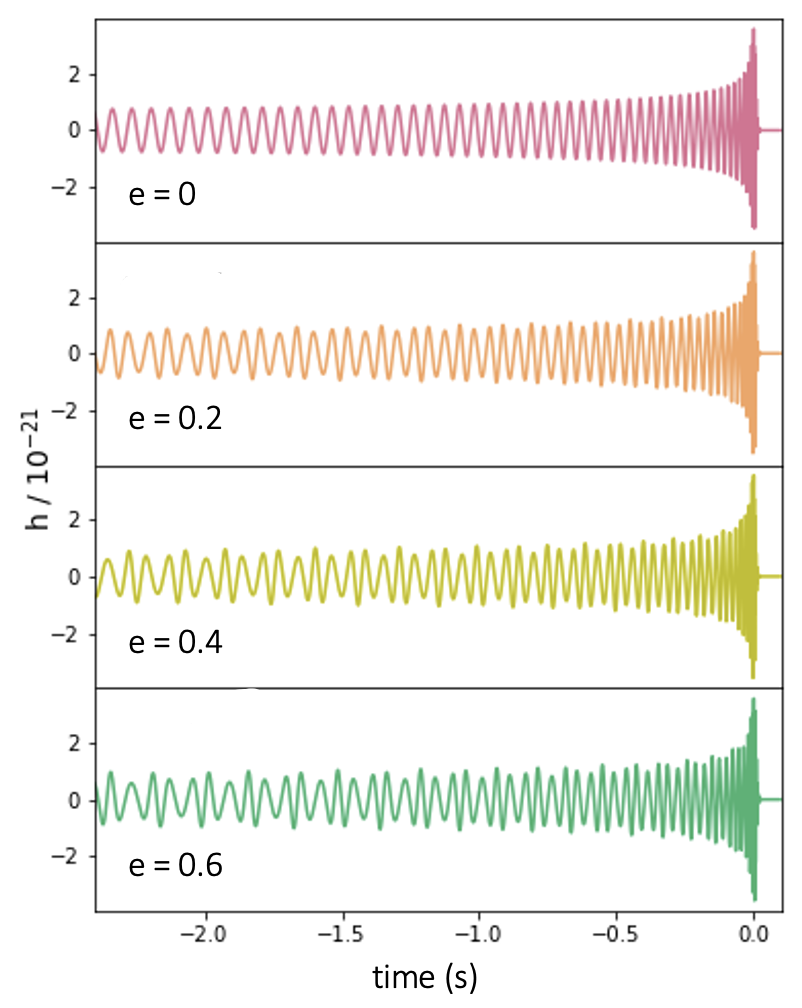}
\caption{ Example strain time series' of $30\,\mathrm{M}_{\odot}$ binary black hole mergers with four different eccentricities calculated at a reference frequency of 10\,Hz. At relatively large eccentricity, modulations of the strain at periapsis can resemble short-lived burst signals, especially at relatively large masses where only a couple of cycles may be present in the LIGO-Virgo-KAGRA observing band. 
}
\label{fig:ebbh}
\end{figure}

Eccentric inspiralling systems circularise due to the emission of gravitational waves~\citep{peters_64}. Systems born through binary stellar evolution are therefore expected to lose all detectable eccentricity by the time they enter the LIGO-Virgo-KAGRA observing band~\citep[e.g.,][]{hinder08}. However, there are two main ways in which a black hole binary system could still be eccentric. First, if the binary system formed dynamically in a dense stellar environment such as a globular cluster or AGN disk~\citep{morscher15,samsing18,rodriguez18}. Second, from hierarchical field triples~\citep{silsbee17,antonini17} and quadruples~\citep{liu19,fragione19} in which the inner binary's eccentricity is driven up through perturbations from the outer component through Kozai-Lidov resonances~\citep{kozai_62,lidov_62}.  For high eccentricities $e\gtrsim0.5$ (see Fig.~\ref{fig:ebbh}) and moderately large masses, modulations of the gravitational-wave strain when the binary is at periapsis imply the signal resembles a short-lived burst immediately prior to merger. The confident detection of such bursts may therefore allow us to distinguish between different formation channels~\citep[e.g.,][]{lower_18}.

Current standard matched filter searches for gravitational waves from binary black holes do not include the effects of eccentricity in their template banks \citep{gwtc1, gwtc2, gwtc3}. This is mainly due to a lack of available waveforms that include the effects of eccentricity, although there are a small number of waveforms currently available \citep{east_13,gayathri22}. There is also a significant increase in computational expense when more waveform parameters are added to the template bank. Templates without eccentricity are still effective at discovering black hole binaries with low ($\lesssim0.2$) eccentricity \citep[e.g.,][]{brown_10,zevin_21}. 

For binary black holes with larger eccentricities ($\gtrsim0.2$), burst-search techniques are required to discover their gravitational-wave signals (although see e.g., ~\citet{calderonbustillo21, gayathri22}). Current LIGO-Virgo-KAGRA searches for gravitational waves from eccentric binaries use the unmodelled burst search algorithm cWB \citep{klimenko_16}, however they have not identified any bona fide eccentric candidates~\citep{eBBH_O1O2, eBBH_O3}.

\begin{figure*}
\includegraphics[width=\textwidth]{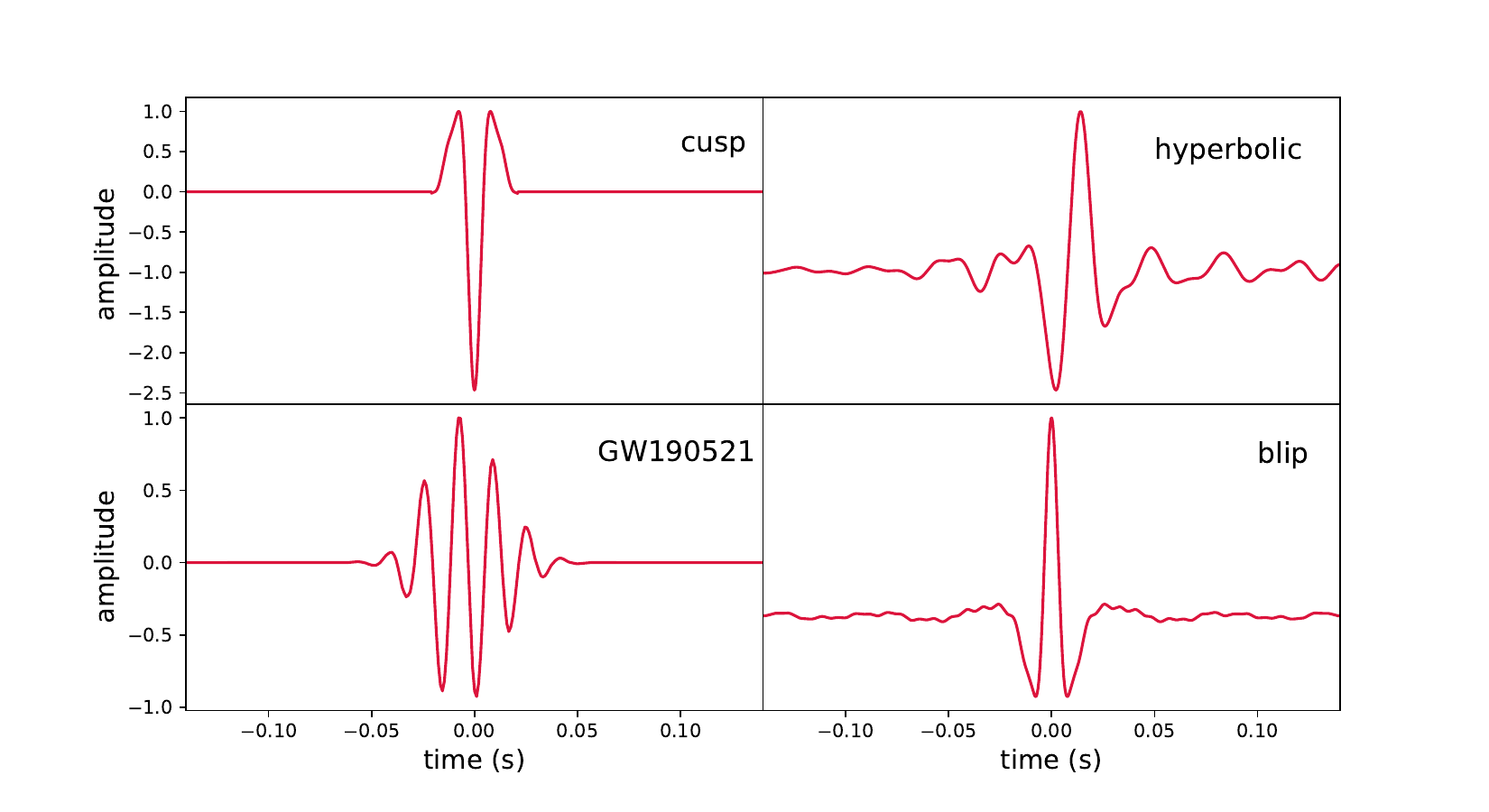}
\caption{Examples of gravitational-wave bursts from different sources. Top left, a cosmic string cusp. Top right, a hyperbolic encounter. Bottom left, a numerical-relativity simulation of the high-mass binary black hole signal GW190521. Bottom right, a blip glitch. Detecting very short duration astrophysical signals is difficult due to their similarity to short duration detector noise glitches.}
\label{fig:string}
\end{figure*}

Black hole systems may also produce short bursts of gravitational waves from hyperbolic encounters~\citep{capozziello_08, devittori_12}. A hyperbolic encounter occurs when black holes in dense star clusters scatter off of one another. If the two black holes pass close enough, then a burst of gravitational waves should occur in the frequency band of current gravitational-wave observatories. The gravitational-wave signal is very short duration, resulting in a single spike in the time series~\citep{bae_20}, as shown in Figure \ref{fig:string}. 

To date, two gravitational-wave searches have been performed that specifically target hyperbolic encounters. \citet{morras_22} searched fifteen days of data during the second LIGO-Virgo-KAGRA observing run, and \citet{bini_24} searched for hyperbolic encounters during O3b. Neither search detected gravitational waves from hyperbolic encounters. \citet{bini_24}, and \citet{dandapat_23}, explored the prospects for detecting these events in future gravitational-wave observing runs. They predict that the fifth observing run of LIGO-Virgo-KAGRA may be sensitivity to hyperbolic encounters out to a volume of $1.33 \pm 0.052 \times 10^{7} \mathrm{Mpc}^{3}$ year.

\section{fast radio bursts}
\label{sec:frb}

FRBs are milli-second duration bursts of radio waves that are often detected at cosmological distances (see \cite{petroff_22, bailes_22} for  recent reviews). The first FRB was discovered by the Parkes/Murriyang telescope \citep{Lorimer_07}, with over a thousand more FRBs discovered to date \citep{petroff_16, chime_21}. Some FRBs have been found to repeat, and a few have been localised to their host galaxies \citep{Spitler_16, chatterjee_17, kumar_19}. 

The astrophysical progenitor of the majority of FRBs is currently unknown. There are a significant number of publications predicting different possible origins of repeating and non-repeating FRBs, some of which predict gravitational-wave counterparts \citep[e.g., see][]{platts_19}. The short time scales of FRBs suggest their origin is some kind of compact object such as a black hole or neutron star, possibly with a strong magnetic field or rapid rotation. However, the FRB detection rates are too high for every FRB progenitor to be a binary neutron star merger \citep{wang_24}.

Potential sources of FRBs could be cataclysmic, although this is likely not the case for repeating FRBs. Examples of cataclysmic potential origins of FRBs are neutron star mergers, extreme supernovae, or neutron star collapse to a black hole. Examples of non-cataclysmic sources include magnetar flares or magnetars in binary systems. Recently, a detection was made of an FRB associated with a galactic magnetar that was found during an X-ray outburst, showing that at least some FRBs originate from magnetars \citep{chime_20}. This event occurred between the current gravitational-wave observatories' third and fourth observing runs. For magnetars to be the origin of extragalactic repeating FRBs, they would need to have a radio luminosity significantly brighter than what we observe from magnetars in our own galaxy.

\begin{figure*}
\centering
\includegraphics[width=12cm]{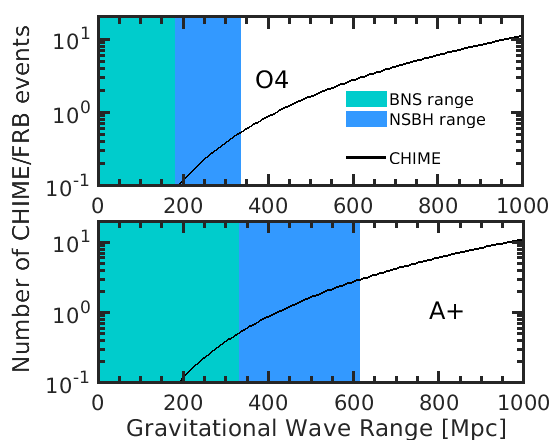}
\caption{The predicted number of future FRB observations, calculated assuming FRBs track the star formation rate, expected to occur within the binary neutron star detection range of the upcoming observing runs of the LIGO-Virgo-KAGRA observatories. A non-detection could rule out a binary neutron star as the FRB progenitor. This figure was produced by Eric Howell, assuming a CHIME detection rate of 2 FRBs per day, and masses of $1.4\,\mathrm{M}_{\odot}$ and $10\,\mathrm{M}_{\odot}$ for neutron star black hole (NSBH) binaries, and masses of $1.4\,\mathrm{M}_{\odot}$ and $1.4\,\mathrm{M}_{\odot}$ for binary neutron stars (BNS).}
\label{fig:frbs}
\end{figure*}

As binary neutron star mergers are one of the proposed progenitors for non-repeating FRBs, searches for their gravitational-wave counterparts use both modelled matched filter compact binary searches and unmodelled gravitational-wave burst search techniques. On the other hand, current searches for repeating FRBs use only unmodelled burst search methods, as it is not possible for a binary neutron star merger to be the progenitor. If an FRB occurs within the gravitational-wave observatories' binary neutron star detection range, then a non-detection could rule out neutron star mergers as the origin. Several FRBs are expected to occur within such a detection range at the expected sensitivity level of the current observatories fifth observing runs, as shown in Figure \ref{fig:frbs}. 

Fast-radio bursts created by magnetar flares are unlikely to be detected in gravitational waves beyond our galaxy. We explore this more in Section~\ref{sec:magnetar}.

To date, there has been only one claim of a coincident gravitational-wave signal with an FRB; that of GW190425 and the CHIME-detected FRB 20190425A~\citep{moroianu23}. The originally-claimed association has a relatively low probability of coincidence of 2.8$\sigma$. However, the association has at least four problems. First, it relies on a long-lived supramassive neutron star surviving 2.5 hours, which requires an incredibly stiff nuclear equation of state given the total inferred mass of the progenitor~\citep{magnahernandez24}. Second, the 400 MHz radio signal is unlikely to traverse the merger ejecta without significant attenuation~\citep{bhardwaj23}. Third, if the association is true then the gravitational-wave data strongly prefer an off-axis system, whereas the FRB requires an on-axis system to be detectable~\citep{bhardwaj23}. Fourth, based on the most likely host galaxy~\citep{panther23,bhardwaj23} and a proper common-source odds statistic~\citep{ashton18}, the significance of the FRB and gravitational-wave event is significantly reduced~\citep{magnahernandez24}.

LIGO-Virgo-KAGRA and GEO searches have not discovered any gravitational-wave counterparts to FRBs to date~\citep{lvk_radio_16, lvk_frb_03a}. The first search used only the GEO and Virgo gravitational-wave observatories, and searched for gravitational-wave burst emission $\pm 2$ minutes around the time of FRBs discovered by the Parkes/Murriyang telescope \citep{lvk_radio_16}. Searches for gravitational-wave counterparts to FRBs discovered by CHIME~\citep{chime_18} started during the third observing run~\citep{lvk_frb_03a}. They searched for 
gravitational waves from the closest 22 non-repeaters, and for eleven repeating bursts from the closest three repeating sources. The search for binary neutron stars covered gravitational-wave observatory data from 10\,s before and 2\,s after the FRB, and the unmodelled search covered 600\,s before the FRB and 120\,s after. This is to ensure that the search window would cover the expected time frame for emission from all theoretical FRB progenitors. Although the previous searches were not able to rule out any possible progenitor sources, this is expected to soon be the case, as the gravitational-wave observatories increase in sensitivity during their next observing runs.

\section{pulsar glitches}
\label{sec:glitch}

Pulsar glitches are sudden increases in the 
angular momentum of the crust of a neutron star causing an inferred increase in the stellar spin frequency~\citep{radhakrishnan69,reichley69,boynton69}. The glitch mechanism is believed to be related to internal superfluid dynamics~\citep{anderson_75}. Broadly speaking, the internal superfluid does not lose angular momentum through the usual spin down process as the superfluid vortices are pinned to lattice sites in the crust; the pinning implies the vortices cannot move outward, and hence the superfluid angular momentum does not change on secular timescales. A lag between the superfluid core and the crystalline lattice in the crust builds up until it is eventually released with angular momentum transferring back from the core to the crust, which is seen as a pulsar glitch.  For a more detailed review of glitch dynamics, see~\citet{haskell_15}, and for associated gravitational-wave emission, see~\citet{haskell24}.

There are two, maybe three, primary mechanisms during a glitch that could generate substantial gravitational waves. First, the spin-up event is expected to ring up multiple large-scale oscillation modes across a broad range of frequencies that would be relatively short lived. Second, relaxation of the stellar interior following the glitch is expected to generate interior bulk-motion fluid flows that can generate gravitational waves through the current quadrupole. And third, but perhaps more speculatively, recent evidence suggests starquakes may accompany glitches~\citep{bransgrove20}, which could also generate gravitational waves. We discuss each of these potential mechanisms presently.

While it is generally believed glitches are triggered by a lag between the superfluid core and the crust, the exact mechanism that triggers the angular momentum transfer from the core is ill-understood. Regardless, recent pulse-to-pulse observations of the Vela 2016 glitch~\citep{palfreyman18} show that the glitch rise time is less than 12\,s (at 90\% confidence), although it could be substantially shorter~\citep{ashton19}. If the timescale is short enough there could be efficient energy conversion between the trigger mechanism and subsequent fluid flow and the fundamental $f$ mode and higher-order $p$ modes~\citep{kokkotas01,sedrakian03,sidery10,ho20,wilson24}, which both couple to the gravitational-wave channel. These modes are short lived, typically damping on timescales $\lesssim0.1$\,s~\citep{detweiler75,lindblom83,mcdermott88}. The sudden glitch rise could also generate inertial $r$ modes~\citep{santiagoprieto12} that are unstable to the Chandrasekhar-Friedmann-Schutz~\citep{chandrasekhar70,friedman78} mechanism, even at low rotation rates~\citep{andersson01} and hence could grow to produce significant gravitational-wave emission. Finally, the spatially inhomogeneous vortex unpinning could generate a current-quadrupole gravitational-wave signal, potentially also observable with current or next-generation observatories~\citep{warszawski_12}. 

While starquake-induced glitch mechanisms are generally disfavoured, the first pulse-to-pulse observation of a glitch~\citep{palfreyman18} showed evidence for a null at the time of the glitch. This is difficult to interpret without significant crustal motion or cracking~\citep{bransgrove20}, which could generate a spectrum of core oscillation modes. Current estimates, however, indicate these could be too small for detection~\citep{keer15,layek20}.

Immediately following a glitch, the stellar core may be rotating slower than the crust, which could lead to non-axisymmetric Eckman flows that could generate gravitational waves through the current quadrupole~\citep{vaneysden08,bennett10,singh17}. Alternatively, glitches could induce neutron star mountains that could generate nearly-chromatic gravitational waves as the star spins~\citep{yim_20,moragues23}, excess superfluid energy could drive gravitational-wave emission~\citep{prix11}, or trapped ejecta mass could cause emission at once and twice the star's spin frequency~\citep{yim24}.

Despite the oodles of proposed mechanisms for generating gravitational waves during and immediately following a neutron star glitch, no such gravitational waves have been observed~\citep{lvk_vela_11,abbott_22_narrowband,lvk_allsky_O3,lopez24}. Some glitch models are already being constrained by these non-observations~\citep[e.g.,][]{yim24b}, implying a future detection of gravitational waves from a glitch may be just around the corner.

\section{magnetars}
\label{sec:magnetar}

Highly magnetised ($B\gtrsim\unit[10^{13}]{G}$) neutron stars---magnetars---exhibit regular bursting events emitting luminosities up to $\approx\unit[10^{43}]{erg\,s^{-1}}$~\cite[e.g., see][for a review]{kaspi_17}, with so-called giant flares emitting peak luminosities up to~$\approx\unit[10^{48}]{erg\,s^{-1}}$. Such extreme events in the stellar magnetosphere may come with crust-cracking, large-scale internal magnetic field rearrangements, and global fluid flows inside the star; all of which are mechanisms proposed to generate gravitational waves. 

Initial estimates for the gravitational-wave energy release proposed that the magnetic field inside the star could be globally rearranged. If the full energy reservoir of this magnetic-field topology change could be converted into gravitational waves as $f$-mode oscillations, then up to $\sim\unit[10^{49}]{erg}$ could be released, making them potentially observable with second-generation observatories~\citep{ioka_01, corsi_11}. Unfortunately, more realistic calculations, both analytic and numerical, tell a more pessimistic story~\citep{levin2011,lasky2011,ciolfi2011,zink2012,tsokaros2022}. These works indicate kHz $f$-mode oscillations will require at least third-generation observatories for galactic giant flares to be detectable.

The $f$-mode oscillations are expected to be short-lived~($\lesssim\unit[100]{ms}$;~\citet{detweiler75,lindblom83,mcdermott88}) and therefore relatively easy to search for (see discussion below). On the other hand, longer lived modes may be excited, but would require detection algorithms that track the phase of the signal. From a theoretical perspective, such longer lived modes include $g$ modes and Alfven waves, where the restoring forces are respectively buoyancy and magnetic fields. Not much theoretical work has been done estimating the potential amplitude, evolution, and damping mechanism of these waves, implying estimates for their total gravitational-wave emission energy and their gravitational-wave detectability are largely unknown. Moreover, searches for such mode oscillations are extremely difficult given the modes are not likely to be exactly chromatic; unknown frequency and phase evolution over long timescales implies fully-coherent search algorithms are not applicable, and there is a significant difference between actual and optimal sensitivity to these signals.

Despite the challenges, the LIGO-Virgo-KAGRA collaboration undertakes searches for gravitational waves from magnetar flares~\citep{lvk_magnetar_07,lvk_magnetar_08,lvk_magnetar_09,lvk_magnetar_11,lvk_magnetar_12,lvk_magnetar_O2,lvk_magnetar_O3}. Of these searches, only one~\citep{lvk_magnetar_07} has been for gravitational waves associated with a giant flare, which was the 2004 giant flare of galactic magnetar SGR 1806-20~\citep{hurley05,palmer05}. That flare was approximately 100 times brighter than the other two giant flares observed, emitting total isotropic energy of $\unit[2\times10^{46}]{erg}$. Importantly, quasi-periodic oscillations were observed in the x-ray tail of the flare~\citep{israel05,watts06,strohmayer06}, which have been attributed to a number of mechanisms associated with oscillation modes in the core, crust, and at the crust-core interface. Despite many of these identified x-ray oscillations having frequencies in the LIGO-Virgo-KAGRA gravitational-wave observing band, no gravitational waves were detected, with upper limits on the gravitational-wave energy emission being $\lesssim\unit[10^{47}]{erg}$. 

The sensitivity of gravitational-wave observatories has greatly improved since the 2004 giant flare, but unfortunately no more galactic giant flares have been observed during this time. We have therefore had to suffice searching data associated with magnetar flares emitting some five or six orders of magnitude less electromagnetic energy~\citep[e.g.,][]{lvk_magnetar_O2,lvk_magnetar_O3}. The best gravitational-wave energy upper limits from these searches are $\lesssim\unit[10^{44}]{erg}$ for both short (e.g., $f$ mode) and long (e.g., $g$ or Alfven mode) searches. For gravitational waves to have been detected from such flares, the gravitational-wave energy emitted from the flare would have had to exceed the electromagnetic energy emitted by some three or four orders of magnitude.

And so we continue to wait for the next giant flare, although there are still no guarantees the network operating at design sensitivity will observe gravitational waves from such bursts.

\section{gamma-ray bursts}
\label{sec:grb}

GRBs are powerful bursts of gamma-rays that are followed by emission in multiple different wavelength bands. There are two classes of GRB, with different progenitors and emission durations. Long GRBs have typical duration larger than 2\,s, and potentially as long as a few hours, and a soft spectral hardness. Short GRBs are dimmer, have duration typically less than 2\,s, and produce hard gamma-rays. 

Long GRBs can last for as long as a few minutes and are thought to be powered by extreme supernova explosions. There is observational evidence for this association through supernovae, such as SN1998bw \citep{galama_98}, which have confirmed long GRB counterparts. Potential progenitors could be the magneto-rotational explosions discussed in Section \ref{sec:ccsn}, which gain some of the energy from the rapid rotation through the magnetic field, or other extreme models such as black hole accretion disks in collapsars \citep{macfadyen_99, gottlieb_24}. 

Short GRBs are generally attributed to compact binary mergers; this was certainly the case for the only multi-messenger gravitational-wave observation that came from the binary neutron star merger GW170817 with an accompanying short GRB~\citep{abbott_17_gw170817_detection,abbott_17_gw170817_multimessenger}. The GRB was several orders of magnitude less energetic than most other GRBs; the combined gravitational-wave and full spectrum of electromagnetic observations suggest the source was detected off-axis. 

Recent electromagnetic observations of some GRBs is now challenging the conventional long/short GRB divide. For example, GRB\,211211A was a typical long GRB with a soft spectrum and burst duration more than 50\,s, albeit with a kilonova counterpart, and therefore strongly suggestive that the progenitor was a compact binary merger~\citep{rastinejad22,troja_22}. That GRB was at a distance of only 350\,Mpc, implying it could have potentially been observed in gravitational waves with LIGO-Virgo-KAGRA fourth observing run sensitivity if it was a binary neutron star merger, and third observing run sensitivity if it was a neutron star-black hole merger~\citep{sarin23,yin23}. The GRB\,230307A was another `typical' long GRB, again with a kilonova; this time, James Webb Space Telescope observations revealed an emission line associated with tellurium, an $r$-process heavy element thought to only be produced through compact binary mergers~\citep{levan24}.

The search for a coincident gravitational-wave and GRB signal from a compact binary merger is much anticipated. That said, in gravitational-wave data analysis parlance, these sources are not considered `burst' gravitational waves because the gravitational-wave signal can be precisely modelled. 

Previous gravitational-wave searches have looked for counterparts to GRBs detected by Fermi and Swift~\citep{GRB_O1, GRB_O2, GRB_O3a, GRB_O3b}. Approximately 100 GRBs occured during 6 months of gravitational-wave observation time, and the majority do not have a known redshift. A model agnostic burst search is used to search for all GRBs regardless of their classification. On the other hand, compact binary searches using template banks of binary inspiral signals were only used to find gravitational-wave counterparts to short GRBs. In general, the gravitational-wave burst searches look for events that occur in a window of 600\,s before and 60\,s after the time of the GRB. It may be worth considering expanding this window given the exotic nature of some recent GRB detections, including those mentioned in previous paragraphs, as well as the recent discovery of soft x-rays preceding gamma-ray observations by several hundred seconds~\citep{liu24}. Previous searches have only searched in the frequency band of $20-500$\,Hz, which is not optimal if the source of the GRB is an extreme supernova because, as described in Section \ref{sec:ccsn}, they are expected to emit most of their gravitational-wave energy at frequencies above 500\,Hz.  

\section{orphan memory}
\label{sec:memory}

Gravitational-wave memory is a non-linear hereditary effect caused by the anisotropic emission of gravitational waves~\citep{zeldovich74,braginsky87,christodoulou91,thorne92}. At frequencies below the characteristic frequency of the gravitational-wave source physics, the gravitational-wave strain amplitude spectral density scales as $S_h^{1/2}\propto1/f$. Any gravitational-wave burst with characteristic frequency above that of the LVK observing band is therefore potentially observable through the low-frequency memory component of the signal~\citep{mcneill_17}.

A number of speculative astrophysical sources have been proposed as emitting gravitational-wave signals at frequencies significantly higher than the $\approx2$ kHz observing band of the LIGO-Virgo-KAGRA observatories. These include cosmic strings, dark-matter collapse in stars, and Kaluza-Klein modes in higher-dimensional theories; for a review, see~\citet{cruise12}. The so-called ``orphan memory'' signals from these high-frequency bursts may be observable in the LIGO-Virgo-KAGRA observing band~\citep{mcneill_17} with strain $S_h^{1/2}\propto1/f$. In such a scenario, distinguishing the source of the memory burst, and even whether it is a memory burst in the first place, could be difficult (although see~\citet{divarkala21}).

The non-existence of loud memory bursts in LIGO-Virgo-KAGRA data can be used to constrain potential signals in dedicated MHz and GHz gravitational-wave observatories; for a review of these observatories see~\citet{aggarwal20}. For example, the Bulk Acoustic Wave High Frequency Gravitational Wave Antenna~\citep{goryachev14} recently announced the detection of ``two strongly significant events''~\citep{goryachev21} that they interpreted \textit{could} have been gravitational waves. However, if those two events are interpreted as gravitational-wave signals, the corresponding memory signals in LIGO/Virgo would have had SNRs in excess of $10^6$~\citep{lasky21}. This \textit{reductio ad absurdum} (together with the cosmological and astronomical implications~\citep{lasky21,domenech21}) show that the signals could not have been gravitational waves, and that the current generations of MHz and GHz observatories are a long way from the required sensitivity to observe real astrophysical signals.

\section{topological defects}
\label{sec:cosmo}

Cosmic strings are topological defects that may have formed during symmetry breaking phase transitions in the early Universe \citep{hindmarsh_95}. Measurements of the cosmic microwave background have shown no evidence for cosmic strings, however gravitational waves may be the best way to provide the first direct evidence for their existence \citep{damour_20}. As well as being a source for gravitational-wave burst searches, a superposition of cosmic strings are also a potential source of a gravitational-wave stochastic background \citep{siemens_07}. Upper limits placed on the emission of gravitational waves from cosmic strings are typically more stringent from stochastic background searches than from gravitational-wave burst searches~\citep{cosmicstring_O3}.  

When a cosmic string crosses itself it can form a loop that becomes detached from the string. The loop will then radiate gravitational waves and eventually disappear \citep{vachaspati_85, casper_95}. A cosmic string cusp is the part of the loop that has detached from the string. The cusps are expected to produce a burst of gravitational waves with an amplitude related to the size of the loop and the tension in the string \citep{stott_17}. An example of a gravitational-wave signal from a cosmic string cusp is shown in Figure \ref{fig:string}. 

Gravitational waves can also be produced by cosmic string kinks, which are localised distortions in the cosmic string that propagate at the speed of light \citep{damour_01}. When two strings cross, they can create a loop that contains several kinks. Gravitational waves from the cusp are emitted in the forward direction of the cusp, and kinks can produce repeating bursts as they move around the cusp. There can also be an isotropic burst of gravitational waves caused by the collision of two kinks on the same cusp. 

The gravitational-wave signals from cosmic strings are understood well enough for a templated matched filter search to be performed. There were no gravitational-wave detections from cosmic strings in the first three LIGO-Virgo-KAGRA observing runs, which searched for cusps, kinks and kink-kink collisions \citep{cosmicstring_O1, cosmicstring_O3}.

The biggest challenge for the detection of cosmic strings is the short duration of the gravitational-wave signal. In the time series, they look very similar to the most common transient noise glitches in gravitational-wave detectors \citep{cabero_19}, which may lead to a large number of false alarms in searches. They also look very similar to the gravitational-wave emission expected from other very short duration astrophysical sources, for example the merger of an intermediate mass black hole binary system. An example of a binary black hole merger identified by a cosmic string search is GW190521, which produced a very short duration gravitational-wave signal in the LIGO-Virgo frequency band due to the high component masses of $85\,\mathrm{M}_{\odot}$ and $66\,\mathrm{M}_{\odot}$ \citep{GW190521_astro}. However, the SNR for the source was much higher in the compact binary search than in the cosmic string search, and Bayesian model selection was also used to rule out a cosmic string origin for this event \citep{GW190521_astro, aurrekoetxea_23}.  

Another potential cosmological source of gravitational-wave bursts is a domain wall passing through a gravitational-wave detector \citep{jaeckel_16}. Quantum field theory tells us that the vacuum is full of fields with an average value of zero. When the Universe expands and cools it results in the production of local minima in the field; domain walls are the boundaries between the areas with different vacuum topology.

The signal morphology and amplitude is dependent on the direction of the domain wall. Due to this, the signal may not have the same morphology in multiple gravitational-wave detectors, implying it could easily be missed by traditional search methods. The duration of the signal is determined by the thickness of the domain wall and the time taken for it to pass through the detector. The speed of the domain wall is typically the speed of the dark matter halo of the galaxy $\sim 300\mathrm{km\,s}^{-1}$.

In gravitational-wave searches, where the travel time of the gravitational-wave is equal to the speed of light, a lot of gravitational-wave transient noise can be rejected because of the time coincidence between the detectors. Because the travel time for a domain wall is much slower than the speed of light, every detector noise glitch is now going to be coincident with other glitches in the data from the other gravitational-wave detectors. This would result in a huge number of false alarms for a domain wall search. Therefore, efforts to reduce detector noise glitches or understand their cause from first principles, are essential to make the first detection of domain walls passing though a gravitational-wave observatory.

\section{unknown unknowns}
\label{sec:unknown}

Every time the Universe has been explored at a new electromagnetic wavelength, new astrophysical objects have been discovered that were not theoretically predicted, and their origins were at least at first unknown. Some examples are GRBs \citep{klebesadel_73} and FRBs \citep{Lorimer_07}. As gravitational-wave observatories become more sensitive, it is likely that they may also begin to detect gravitational waves from sources that have not already been theoretically predicted. This is arguably the most exciting aspect of this burgeoning field.

\begin{figure}
\includegraphics[width=\columnwidth]{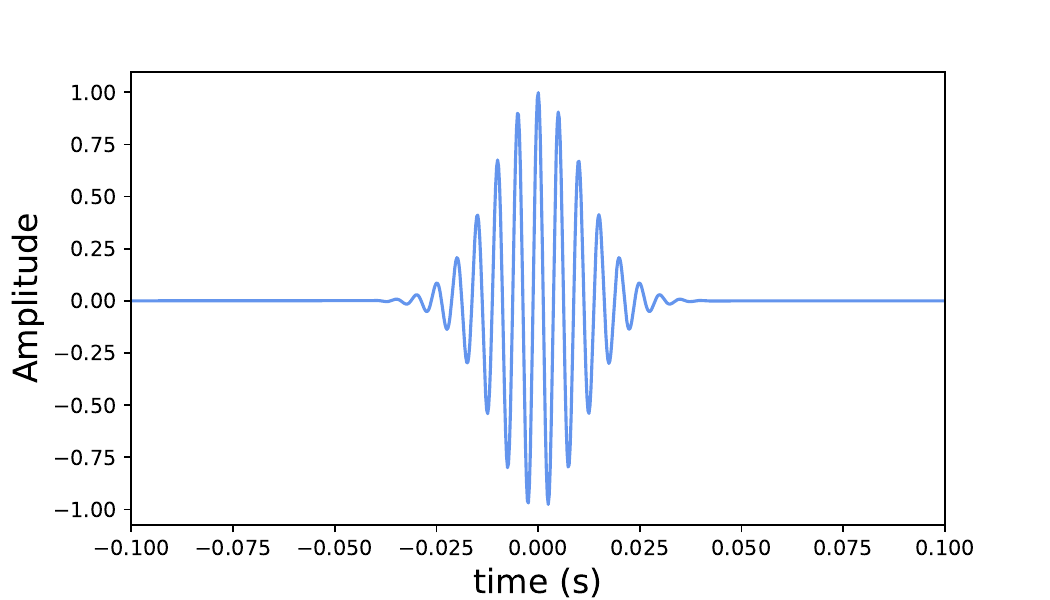}
\caption{A typical sine Gaussian wavelet with a frequency of 200\,Hz, and a duration of 0.05\,s. Sine Gaussian wavelets are often used as signal models for searches and waveform reconstruction of gravitational-wave bursts with an unknown signal shape. 
}
\label{fig:sinegauss}
\end{figure}

Searching for completely unknown gravitational-wave sources presents significant detection challenges. Searches for unknown bursts are divided into short-duration searches of less than a few seconds, and long-duration searches of up to a few hundred seconds. The burst search algorithms  \citep[e.g.,][]{klimenko_16, lynch_17, cornish_15, sutton_10, skliris_20, thrane_11, thrane_15} typically use one or more sine Gaussian, sine Gaussian wavelet, or other common wavelet types as basis functions for their signal model. An example of a typical sine Gaussian is shown in Figure \ref{fig:sinegauss}. These types of generic signal models are good at adapting to fit an unknown shape. The duration and frequency can be altered to cover the burst signal parameter space, with the typical frequency range of such burst searches covering $\sim 30-2000$\,Hz; the most sensitive frequency band of the current observatories.

\begin{figure*}
\includegraphics[width=\textwidth]{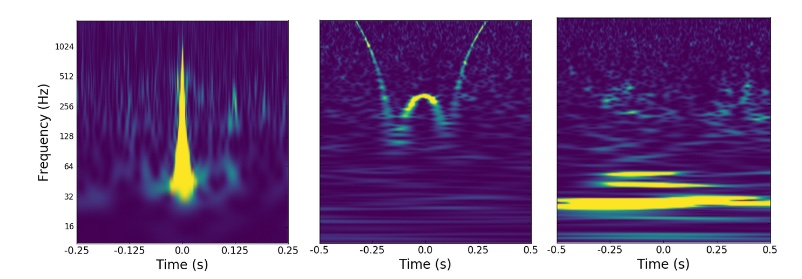}
\caption{ Examples of detector noise glitches from Advanced LIGO. Glitches limit the gravitational-wave search sensitivity, and can contaminate astrophysical signals. From left to right, a blip glitch, a whistle glitch and scattered light. Figure reproduced from \citet{powell_18}. }
\label{fig:glitch}
\end{figure*}

Gravitational-wave data contains many transient noise artifacts, known as glitches, that can mimic gravitational-wave burst signals. Some of the most typical types of glitches that limit the background of gravitational-wave burst searches are shown in Figure \ref{fig:glitch}. Due to these glitches, a coincident detection between multiple interferometers is needed for confidence that a gravitational-wave burst signal has an astrophysical origin. For this reason, gravitational-wave burst searches look for signals with common shape and arrival time (within the light-travel time of the observatories) in multiple detectors \citep{klimenko_16, lynch_17, cornish_15}. They then estimate the gravitational-wave background by time-sliding the data to measure how often glitches accidentally occur at the same time in multiple detectors \citep{was_10}. The imperfect duty cycles imply only one gravitational-wave observatory is operational about 20\% of the time. To ensure we do not miss a gravitational-wave burst during single-detector observing time, better methods for the elimination of glitches are required \citep{cavaglia_20}. To date, the current gravitational-wave all-sky burst searches have only detected gravitational waves from binary black holes  \citep{lvk_allsky_O1, lvk_allsky_O2, lvk_allsky_O3, lvk_long_O1, lvk_long_O2, lvk_long_O3}. For the LIGO-Virgo-KAGRA third observing run, the all-sky short duration burst searches were sensitive to sources with gravitational-wave energy as low as $10^{-10}\mathrm{M}_{\odot}c^{2}$ at a distance of 10\,kpc and a frequency of 70\,Hz. 

The all-sky searches for unknown gravitational-wave bursts can produce parameter estimates for the sky location, the gravitational-wave amplitude, central frequency and duration \citep{essick_15, becsy_17}. They can also reconstruct the waveform of the gravitational-wave signal \citep{millhouse_18, raza_22}. As for compact binary signals, the accuracy of the sky location is largely dependent on the number of gravitational-wave observatories operating. How well the overall signal morphology can be reconstructed depends primarily on the SNR and the complexity of the signal. In general, longer duration bursts and more broadband signals are more difficult to reconstruct. 

Understanding the astrophysical source properties from the parameters measured by the burst searches will be difficult in the case of an unknown source, especially if there is no electromagnetic or neutrino counterpart. 
The observed waveform could, in principle, be used to place limits on the characteristic mass and size of the progenitor system, albeit contingent on the likely unknown distance~\citep{bescy_20}. Such constraints would be based on there being a maximum compactness and a maximum luminosity for any potential gravitational-wave source. If the luminosity distance to the source is known from an electromagnetic counterpart, we could also place constraints on the gravitational-wave energy, and set lower limits on the mass and size of the astrophysical source.

\section{conclusions}
\label{sec:conclusion}

The current gravitational-wave observatories have already transformed our understanding of the Universe, producing the first direct detection of gravitational waves, the first strong field tests of general relativity, and the first direct evidence that binary neutron stars are the progenitors of short GRBs. However, this is just the beginning. The next generation of observatories will hear every binary black hole merger in the observable Universe, and almost all of the binary neutron star mergers well back before the peak of star formation. However, there are many other potential sources of gravitational waves that have yet to be discovered. A detection of a burst of gravitational waves from a different type of source is one of the most promising potential next gravitational-wave discoveries.

In this Review, we describe the expected future sources of bursts of gravitational waves, the prospects for their detection, and what we might learn about the astrophysics of the source. The gravitational-wave signals of some astrophysical burst sources, such as CCSNe, are fairly well modelled, however their exact waveforms are difficult to predict due to complex input physics and stochastic processes. Hydro-dynamical simulations of these systems have allowed us to understand their gravitational-wave signals enough to enable the community to develop tools to extract astrophysical parameters of the source, such as the stars rotation and equation of state. Gravitational-wave signals from highly eccentric compact binaries are also fairly well understood, but not well enough to produce a full template bank to perform a matched filter search.

Gravitational-wave burst signals from other sources are significantly more difficult to model, as not enough is known about their astrophysical progenitors. FRBs currently have many theories for their potential progenitors, which translate to a large range in different predictions for the gravitational-wave signal shape, duration, frequency, and amplitude. It is also possible that there may be sources of gravitational waves that have not been predicted theoretically or observed electromagnetically. A detection of a gravitational-wave burst signal from this type of unknown source would result in a significant advancement in our understanding of the Universe. 

\section{acknowledgments}

We acknowledge helpful comments on this work from Simon Stevenson, Isobel Romero-Shaw, Eric Howell, Sophie Bini and Ik Siong Heng.  
The authors are supported by the Australian Research Council (ARC) Centre of Excellence for Gravitational Wave Discovery OzGrav project numbers CE170100004 and CE230100016.  JP is supported by the ARC Discovery Early Career Researcher Award (DECRA) project number DE210101050. PDL is supported through ARC Discovery Projects DP220101610 and DP230103088, and LIEF Project LE210100002. 


\bibliographystyle{pasa-mnras}
\bibliography{bib}

\end{document}